\documentclass[twoside,twocolumn,9pt]{article}
\usepackage{extsizes}
\usepackage[super,sort&compress,comma]{natbib}
\usepackage[version=3]{mhchem}
\usepackage[left=1.5cm, right=1.5cm, top=1.785cm, bottom=2.0cm]{geometry}
\usepackage{balance}
\usepackage{times,mathptmx}
\usepackage{sectsty}
\usepackage{graphicx}
\usepackage{lastpage}
\usepackage[format=plain,justification=justified,singlelinecheck=false,font={stretch=1.125,small,sf},labelfont=bf,labelsep=space]{caption}
\usepackage{float}
\usepackage{fancyhdr}
\usepackage{fnpos}
\usepackage[english]{babel}
\addto{\captionsenglish}{%
  
}
\usepackage{array}
\usepackage{droidsans}
\usepackage{charter}
\usepackage[T1]{fontenc}
\usepackage[usenames,dvipsnames]{xcolor}
\usepackage{setspace}
\usepackage[compact]{titlesec}
\usepackage{hyperref}

\usepackage{epstopdf}

\definecolor{cream}{RGB}{222,217,201}

\begin{document}

\makeFNbottom
\makeatletter
\renewcommand\LARGE{\@setfontsize\LARGE{15pt}{17}}
\renewcommand\Large{\@setfontsize\Large{12pt}{14}}
\renewcommand\large{\@setfontsize\large{10pt}{12}}
\renewcommand\footnotesize{\@setfontsize\footnotesize{7pt}{10}}
\makeatother

\renewcommand{\thefootnote}{\fnsymbol{footnote}}
\renewcommand\footnoterule{\vspace*{1pt}%
\color{cream}\hrule width 3.5in height 0.4pt \color{black}\vspace*{5pt}}
\setcounter{secnumdepth}{5}

\makeatletter
\renewcommand\@biblabel[1]{#1}
\renewcommand\@makefntext[1]%
{\noindent\makebox[0pt][r]{\@thefnmark\,}#1}
\makeatother
\renewcommand{\figurename}{\small{Fig.}~}
\sectionfont{\sffamily\Large}
\subsectionfont{\normalsize}
\subsubsectionfont{\bf}
\setstretch{1.125} 
\setlength{\skip\footins}{0.8cm}
\setlength{\footnotesep}{0.25cm}
\setlength{\jot}{10pt}
\titlespacing*{\section}{0pt}{4pt}{4pt}
\titlespacing*{\subsection}{0pt}{15pt}{1pt}

\makeatletter
\newlength{\figrulesep}
\setlength{\figrulesep}{0.5\textfloatsep}

\newcommand{\topfigrule}{\vspace*{-1pt}%
\noindent{\color{cream}\rule[-\figrulesep]{\columnwidth}{1.5pt}} }

\newcommand{\botfigrule}{\vspace*{-2pt}%
\noindent{\color{cream}\rule[\figrulesep]{\columnwidth}{1.5pt}} }

\newcommand{\dblfigrule}{\vspace*{-1pt}%
\noindent{\color{cream}\rule[-\figrulesep]{\textwidth}{1.5pt}} }

\makeatother

\twocolumn[
  \begin{@twocolumnfalse}
\vspace{1cm}
\begin{tabular}{m{17cm} p{13.5cm} }
\begin{center}
\noindent\LARGE{\textbf{Flexo-electricity of the dowser texture}} \\
\vspace{1cm}

\noindent\large{Pawel Pieranski$^{\ast}$\textit{$^{a}$} and Maria Helena Godinho\textit{$^{b}$}} \\
\end{center}
\\
\\
\textbf{Abstract:} The persistent quasi-planar nematic texture known also as the dowser texture is characterized by a 2D unitary vector field \textbf{d}. We show here that the dowser texture is sensitive, in first order, to electric fields.  This property is due to the flexo-electric polarisation \textbf{P} collinear with \textbf{d} expected from R.B. Meyer's considerations on flexo-electricity in nematics. It is pointed out that due to the flexo-electric polarisation nematic monopoles can be manipulated by electric fields of appropriated geometry.
\vspace{1cm}
\end{tabular}
\end{@twocolumnfalse}

]

\footnotetext{\textit{$^{a}$~Laboratoire de Physique des Solides, Universit\'{e} Paris-Sud, 91405 Orsay, France. Tel: 33 1 6915 7285; E-mail: pawel.pieranski@u-psud.fr}}
\footnotetext{\textit{$^{b}$~CENIMAT, Faculdade de Ci\^{e}ncias e Tecnologia - Universidade Nova de Lisboa Campus da Caparica, 2829 - 516 Caparica, Portugal. }}

\vspace{2cm}
\section{Introduction}
\label{sec:Intro}
\subsection{The dowser texture and its order parameter}
\label{sec:Dowser_texture}
Generation and handling of monopoles in nematics are two challenging tasks emerging in the field of topological metamaterials. Giomi et al.  \cite{Giomi_Sengupta} proposed recently a micro-fluidical method of handling monopoles. They have shown that nematic monopoles can be trapped in stagnation points of different topological charges created by appropriate microfluidic junction.

Alternatively, monopoles can be generated and handled in the persistent quasi-planar nematic texture, called the dowser texture (see below), which can be considered as a natural universe of nematic monopoles for reasons expounded below \cite{PP_Copar_MHG_hedgehog,PP_Copar_MHG,Pieranski_reflets}.

\begin{figure}
\begin{center}
\includegraphics {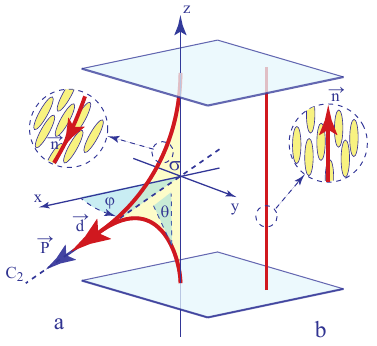}
\caption{Textures of a nematic confined between two surfaces with homeotropic anchoring conditions: a) the dowser (quasi-planar) texture. b) the homeotropic texture.                    \label{fig:Dowser_state}}
\end{center}
\end{figure}

The three-dimensional quasi-planar texture \textbf{n}(z) (see Fig.\ref{fig:Dowser_state}a):
\begin{equation}\label{eq:dowser_field}
\textbf{n}=(\sin\theta\cos\varphi,\sin\theta\sin\varphi,\cos\theta)\ \textrm{with}\  \theta=\pi z/h
\end{equation}
is generated when a gap between two surfaces with homeotropic anchoring conditions  is filled for the first time with a nematic. It is in competition with the distorsionless homeotropic uniform texture \textbf{n}(z)=(0,0,1) which usually, for energetical reasons, eliminates the quasi-planar one.

The symmetry C$_{2v}$ of the quasi-planar texture is lower then the symmetry D$_{\infty h}$ of the homeotropic one. The order parameter, resulting from the symmetry breaking D$_{\infty h}\rightarrow$ C$_{2v}$, is a two dimensional unitary vector field
\begin{equation}\label{eq:dowser_field_d}
\textbf{d}=(\cos\varphi,\sin\varphi)\
\end{equation}
called \emph{the dowser field}, which is fully determined by its azimuthal angle $\varphi (x,y)$ called \emph{phase}. Let us stress that $\textbf{d}\not\equiv-\textbf{d}$, contrary to the nematic director field for which $\textbf{n}\equiv-\textbf{n}$. Alternatively, a complex order parameter $e^{i\varphi}$ can also be used. Let us emphasize that contrary to quantum systems such as superfluids, superconductors, or cold atoms condensates which are also equipped with the complex order parameter $\Psi e^{i\varphi}$, the phase $\varphi$ of the dowser texture has a direct geometrical meaning (see Fig. \ref{fig:Dowser_state}).
\subsection{Metastability of the dowser texture}
\label{sec:Dowser_texture_metastability}
For decades, the quasi-planar texture has been considered as ephemeral because of its metastabilty with respect to the undistorted homeotropic ground state. For this reason it has been scarcely studied in past. The breakthrough came from the discovery made for the first time by Gilli \emph{et al.} \cite{Gilli} and more recently in ref.\cite{PP_Copar_MHG} that in thick enough samples, limited laterally by an air/nematic interface with homeotropic boundary conditions (see Fig.\ref{fig:Setup}a), the homeotropic texture can be eliminated for the benefit of the quasi-planar one. Upon a subsequent reduction of the sample thickness, the metastable quasi-planar texture persists until the thickness h is reduced below $\approx 1\mu m$. Because of its resemblance with a wooden dowser tool such a long-lived quasi-planar texture has been dubbed "the dowser texture".
\begin{figure*}
\begin{center}
\includegraphics {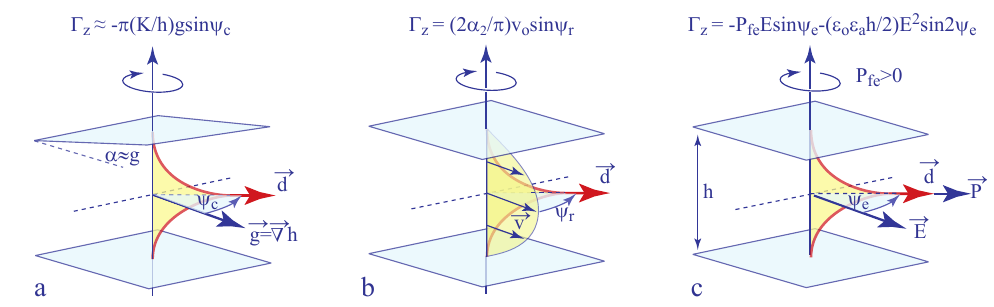}
\caption{Tropisms of the dowser texture. a) Cuneitropism: orientation of the dowser field \textbf{d} by a thickness gradient \textbf{g}. b) Rheotropism: orientation of the dowser field \textbf{d} by a Poiseuille flow \textbf{v}. c) Electrotropism: orientation of the dowser field by an electric field. It is due to the torque exerted by the electric field \textbf{E} on the polarisation $\textbf{P}=P_{fe}\textbf{d}+(\epsilon_{o}\epsilon_{a}h/2)(\textbf{E}\cdot\textbf{d})\textbf{d}$ of the dowser texture. It has two contributions: the spontaneous flexo-electric polarisation and the anisotropic part of the induced polarisation.        \label{fig:Tropisms}}
\end{center}
\end{figure*}
\subsection{Properties of the dowser texture}
\label{sec:Dowser_texture}
The discovery of the method allowing to preserve indefinitely the dowser texture allowed to unveil its properties:
\begin{description}
  \item[The dowser texture as an universe of nematic monopoles \cite{Pieranski_reflets}:] The $+2\pi$ and $-2\pi$ point defects of the 2D dowser field \textbf{d}, when considered in three dimensions of the nematic layer, appear to be nothing else but point defects of the 3D director field \textbf{n}. We call them respectively \emph{monopole}, for the $+2\pi$ defect, and \emph{antimonopole} for the $-2\pi$ defect \cite{PP_Copar_MHG,PP_Copar_MHG_hedgehog}.
  \item[Tropisms of the dowser  texture:] Due its degeneracy with respect to the phase angle $\varphi$, the dowser field is sensitive to perturbations such as a thickness gradient  \textbf{grad}h or a Poiseuille flow \textbf{v}. As the dowser field \textbf{d} tends to align in directions of the thickness gradient and of the flow we called these properties respectively \emph{cuneitropism}  \cite{PP_Copar_MHG} and \emph{rheotropism} \cite{PP_Hulin_MHG_rheo} (see Fig.\ref{fig:Tropisms}a and b).
\end{description}
\subsection{Electrotropism of the dowser texture}
\label{sec:Electrotropism}
In the present work we deal with \emph{electrotropism} corresponding to the alignment of the dowser field by electric fields (see Fig.\ref{fig:Tropisms}c). It is due to the torque (per unit area)
\begin{equation}\label{eq:torque_pol}
   \overrightarrow{\Gamma}_{e}=\overrightarrow{P}\times \overrightarrow{E}
\end{equation}
exerted by the electric field, parallel to the xy plane of the nematic layer, on the polarisation \textbf{P} being a sum
\begin{equation}\label{eq:polarisation}
   \overrightarrow{P}=\overrightarrow{P_{fe}} + \overrightarrow{P_{ind}}(\vec{E})
\end{equation}
of the spontaneous flexo-electric polarisation (per unit area, see the next section)
\begin{equation}\label{eq:polarisation_flexo}
\overrightarrow{P_{fe}}=P_{fe}\vec{d}
\end{equation}
and of the anisotropic part of the polarization  $\overrightarrow{P_{ind}}$ (per unit area) induced by the electric field:
\begin{equation}\label{eq:polarisation_induced}
   \overrightarrow{P_{ind}}=\epsilon_{o}\epsilon_{a}h(\vec{d}\cdot\vec{E})\vec{d}
\end{equation}
When the electric field makes with the dowser field an angle $\psi_{e}$ (see Fig.\ref{fig:Tropisms}c), the electric torque given by equation \ref{eq:torque_pol} can be written as:
\begin{equation}\label{eq:torque_pol_2}
   \Gamma_{ez}=-P_{fe}E\sin\psi_{e}-(\epsilon_{o}\epsilon_{a}h/2)E^{2}\sin2\psi_{e}
\end{equation}
The relative importance of the two terms depends on the intensity of the electric field E with respect to a characteristic field E$_{c}$ defined as
\begin{equation}\label{eq:Ec}
E_{c}=\frac{P_{fe}}{\epsilon_{o}\epsilon_{a}h/2}
\end{equation}
\begin{figure*}
\begin{center}
\includegraphics{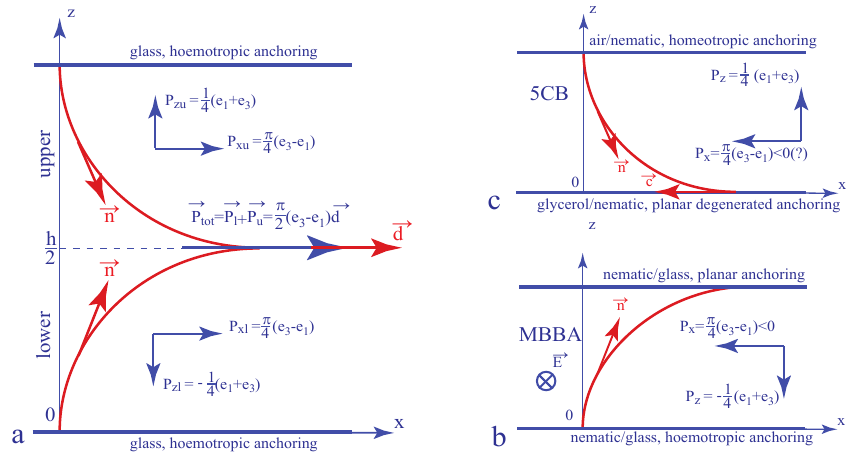}\\
\caption{Flexo-electric polarisation in the dowser texture compared with that of nematic layers with hybrid anchoring conditions. a) The dowser texture studied here. b) Experiment of Dozov \emph{et al.} \cite{Dozov} with MBBA, $e_{3}-e_{1}<0$. c) Experiment of Link \emph{et al.}\cite{Takezoe} with 5CB, $e_{3}-e_{1}<0$ (?).  }
\label{fig:Flexo_pol}
\end{center}
\end{figure*}
\subsection{Flexo-electric polarization of the dowser texture}
\label{sec:Polarisation}

One of the principal results of this work is the detection, for the first time, of the flexo-electric polarisation \textbf{P}$_{fe}$ of the dowser texture which is expected from symmetry arguments enounced for the first time by of R.B. Meyer \cite{Meyer}. In the present case, the C$_{2v}$ symmetry allows the dowser texture to have a polarisation per unit area directed along the C$_{2}$ symmetry axis parallel to \textbf{d}: \textbf{P}$_{fe}$=P$_{fe}$\textbf{d}.

It can be calculated by integration of the flexo-electric polarisation density:
\begin{equation}\label{eq:mean_polarisation_1}
\textbf{P}_{fe}=\int_{0}^{h}(e_{1}\mathbf{n}\overrightarrow{\nabla}\cdot\mathbf{n}+e_{3}(\overrightarrow{\nabla}\times\mathbf{n})\times\mathbf{n})dz
\end{equation}
with coefficients e$_{1}$ and e$_{3}$ corresponding to the de Gennes' definition \cite{PGG}. Using equation \ref{eq:dowser_field}, one obtains:
\begin{equation}\label{eq:mean_polarisation_2}
\textbf{P}_{fe}=\mathrm{P}_{fe}\textbf{d}\ \textrm{with} \ \mathrm{P}_{fe}=\frac{\pi}{2}(e_{3}-e_{1})
\end{equation}
\subsection{Former works on flexo-electricity of a nematic layer with hybrid anchoring}
\label{sec:Polarisation}
The flexo-electric polarization per unit area of the dowser texture is related to that of a nematic layer with hybrid anchoring conditions - planar on one limit surface and homeotropic on the other one - which has been studied extensively in past. Indeed, as shown in Fig.\ref{fig:Flexo_pol}, the dowser texture is a superposition of two hybrid aligned textures - a lower one (labelled $l$) for 0$<$z$<$h/2  and upper one (labelled $u$) for h/2$<$z$<$h. The two hybrid aligned textures are related by the C$_{2}$ symmetry of the dowser texture.

When considered separately, these hybrid aligned textures have polarisations per unit area given by:
\begin{eqnarray}\label{eq:mean_polarisation_2}
\textbf{P}_{l} &=& \left(\frac{\pi}{4}(e_{3}-e_{1}),0,-\frac{1}{4}(e_{1}+e_{3})\right)\\
\textbf{P}_{u} &=& \left(\frac{\pi}{4}(e_{3}-e_{1}),0,+\frac{1}{4}(e_{1}+e_{3})\right)
\end{eqnarray}
Let us emphasize that signs of the x and z components of the flexo-electric polarisations displayed in Fig.\ref{fig:Flexo_pol}a correspond to e$_{3}$-e$_{1}>$0 and e$_{1}$+e$_{3}>$0.

First experiments with the hybrid geometry were made by I. Dozov \emph{et al.} \cite{Dozov}. The electric field \textbf{E} applied in y-direction exerted a $\Gamma_{z}$ torque on the P$_{xl}$ component of the polarisation. Let us stress that in this hybrid geometry (identical with lower half of the dowser texture in Fig.\ref{fig:Flexo_pol}), the planar anchoring on the upper glass slide hinders rotation of the splay-bend texture so that its deformation is torsion-like. The value of the pertinent flexo-electric coefficient obtained by Dozov \emph{et al.} with MBBA was
\begin{equation}\label{eq:Dozov}
 e_{3}-e_{1}=-3.3 pC/m
\end{equation}
Its negative sign means that the orientation of the flexo-electric polarisation \textbf{P}$_{xl}$ in MBBA is contrary to the one in Fig.\ref{fig:Flexo_pol}a.

More recently, D.R. Link et al. performed experiments with a 5CB nematic film floating on the surface of glycerol \cite{Takezoe} where the splay-bend texture (identical with upper half of the dowser texture in Fig.\ref{fig:Flexo_pol}) is free to rotate around the z axis similarly to the dowser texture in our experiments discussed below. The local azimuthal orientation of this floating splay-bend texture was indicated by a 2D unitary vector field \textbf{c} similar to the dowser field \textbf{d} except for its sign which was taken opposite to \textbf{d}. Using this convention, the in-plane component of the flexo-electric polarisation can be written as:
\begin{equation}\label{eq:pol_Link}
\textbf{P}_{x}=-\frac{\pi}{4}(e_{3}-e_{1})\textbf{c}
\end{equation}
This floating, free-to-rotate splay-bend hybrid texture was submitted to an in-plane electric field. From the width of 2$\pi$ walls induced by the electric field \textbf{E} in the \textbf{c} field (see section \ref{sec:walls_wound_up}), Link \emph{et al.} determined the value of the pertinent flexo-electric coefficient in 5CB :
\begin{equation}\label{eq:Dozov}
 e_{3}-e_{1}=-11 pC/m
\end{equation}
This means that the in-plane flexo-electric polarisation \textbf{P}$_{x}$ detected by by Link \emph{et al.} in 5CB would have the same direction as the one in MBBA: being parallel to \textbf{c} it would be opposite to \textbf{d}.
\begin{figure*}
\begin{center}
\includegraphics {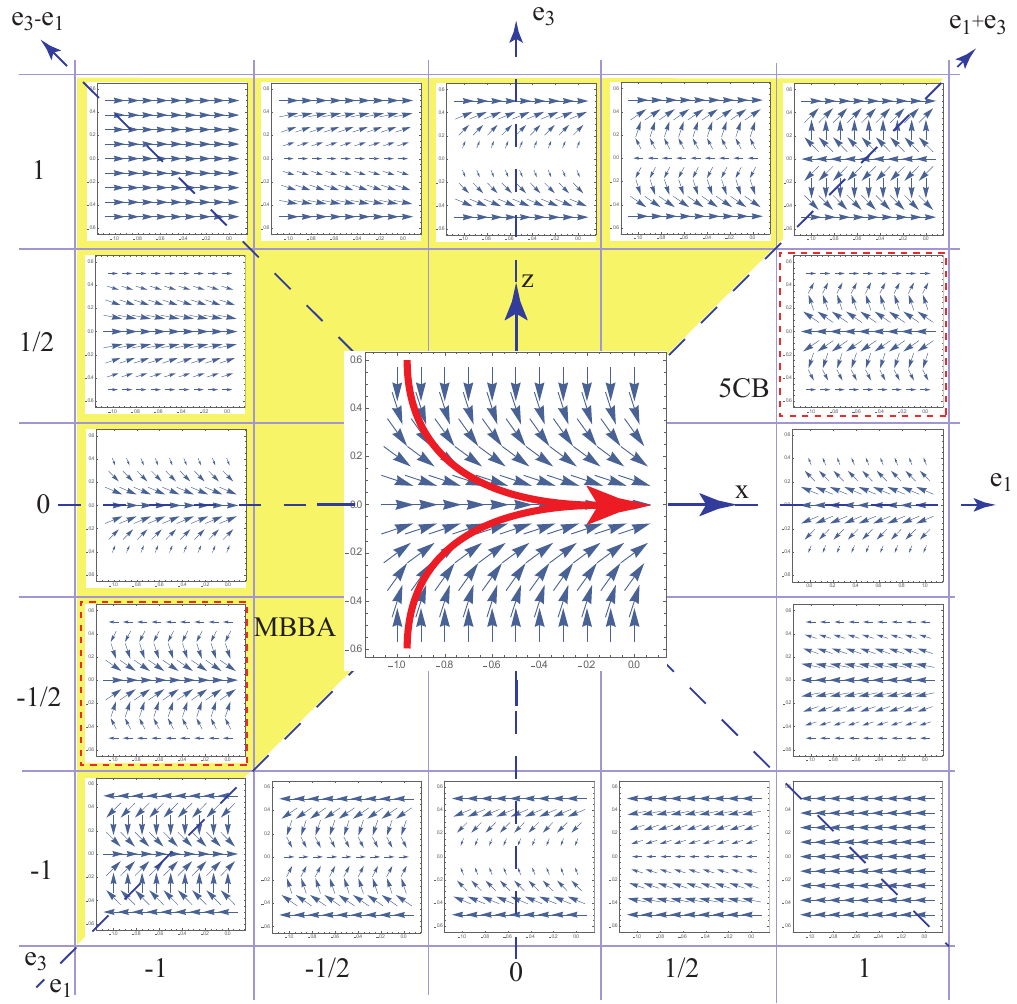}
\caption{Typical aspects of the flexo-electric polarisation density $\textbf{P}(x,y)$ in the dowser texture. The director field $\textbf{n}(x,y)$ of the dowser texture given by eq.\ref{eq:dowser_field} is represented in the central picture. The surrounding 16 polarisations patterns have been calculated for discrete values of the flexo-electric coefficients defined by $(e1,e3)=(i/2,j/2)$ with $-2<i<2$ and $-2<j<2$. The two patterns surrounded by dashed lines and labeled MBBA and 5CB are discussed in section \ref{sec:qual_expl}. }
\label{fig:polar_e1_e3}
\end{center}
\end{figure*}

As we will see below, our experiments with the dowser texture are similar both to those of Link \emph{et al.} and of Dozov et al. because the electric field was applied in the xy plane of the nematic layer. In other experiments with the hybrid splay-bend texture, the electric field was applied in z direction orthogonal to the nematic layer. For this reason we will not discuss them here.
\subsection{The aim of the present work}
\label{sec:Polarisation}
Stimulated by former experiments of Giomi \emph{et al}.\cite{Giomi_Sengupta} on trapping of antimonopoles in stagnation points of flows in microfluidic junctions, we plan to use the expected flexoelectric polarisation of the dowser texture for manipulation of defects of the dowser field by electric fields of appropriate geometry.

Here we report only on preliminary experiments performed with the aim to detect the sign and the amplitude of the flexo-electric polarisation in MBBA and 5CB. For this purpose we used the simplest geometry of a homogeneous electric field applied in the direction orthogonal to the dowser field.

As we will see below, in the case of MBBA our result, \textbf{P} antiparallel to \textbf{d} (\emph{i.e.} $e_{3}-e_{1}<0$), agrees in sign with findings of Dozov \emph{et al.} however, in the case of 5CB, we have found, contrary to Link \emph{et al.} that \textbf{P} is parallel to \textbf{d} (\emph{i.e.} $e_{3}-e_{1}>0$).
\begin{figure}
\begin{center}
\includegraphics{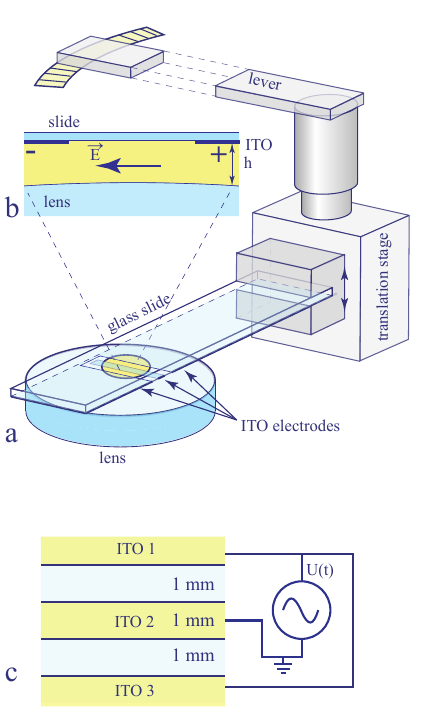}\\
\caption{Setup: a) general perspective view, b) cross section of the nematic droplet squeezed between the lens and the glass slide equipped with ITO electrodes located at its lower surface, c) crankshaft coupled to a stepping motor and mounted on a translation stage, c) systems of ITO electrodes. }
\label{fig:Setup}
\end{center}
\end{figure}

As we have found it quite easy to make a sign error during determination of the pertinent coefficient $e_{3}-e_{1}$, we represented in Fig.\ref{fig:polar_e1_e3} 16 typical aspects of the flexo-electric polarisation density calculated for the dowser texture, defined by Eq.\ref{eq:dowser_field}, using 16 sets of flexo-electric ceoffcients defined by $(e_{1},e_{3})=(i/2,j/2)$ with $-2<i<2$ and $-2<j<2$.

This scheme shows clearly that the mean polarisation $\textbf{P}=\int_{0}^{h}\textbf{P}(x,z)dz$ is parallel (antiparallel) to \textbf{d} when $e_{3}-e_{1}$ is positive (negative).

\section{Experimental}
\subsection{Setup}
For the study of the interaction of the dowser texture with electric fields and, in particular, for the detection of the flexo-electric polarisation of the dowser texture we used the setup depicted in Fig.\ref{fig:Setup}. It is identical with the one that has been described in more details in ref.\cite{PP_Hulin_MHG_rheo} except for the glass slide which now can be equipped with ITO electrodes of various geometries.

In the work presented here we used the two-gap system of three ITO electrodes depicted in Fig.\ref{fig:Setup}c. The central electrode ITO2 is grounded, while the two other electrodes ITO1 and ITO3 are connected together and fed with an AC alternating voltage U(t)=U$_{o}\cos(2\pi ft)$ of frequency f=20mHz. The amplitude U$_{o}$ was 5V in the case of MBBA and 3V in the case of 5CB. With the width of the gap l$_{g}$=1mm one obtains thus E$_{max}$=5 or 3 V/mm.

Each experiment was made in three steps:
\begin{enumerate}
  \item Generation of the dowser texture by an adequate manipulation of the sample thickness (see ref.\cite{PP_Copar_MHG_hedgehog}).
  \item Relaxation of the dowser field to its radial configuration by action of the cuneitropism.
  \item Application of electric fields by means of systems of ITO electrodes located at the lower surface of the glass slide.
\end{enumerate}
\subsection{Conductive and dielectric regimes}
\label{sec:radial_dowser_field}
The geometry of the electric field generated by our systems of ITO electrodes depends crucially on the frequency f of the potential difference U(t)=U$_{o}$cos(2$\pi$ft) applied to them. Indeed, it is well known since the first studies on electrohydrodynamic instabilities in nematics that two different regimes are possible (see f.ex. ref.\cite{PGG}).
\begin{figure}
\begin{center}
\includegraphics{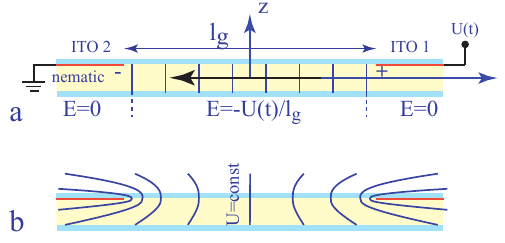}\\
\caption{Generation of the electric field in the conductive and dielectric regimes. a) Equipotentials in the low frequency conductive regime. b) Equipotentials in the high frequency dielectric regime.}
\label{fig:cond_diel}
\end{center}
\end{figure}
Let $\tau_{c}=RC$ be the characteristic time of charges relaxation in our sample considered as a capacitor C, formed by ITO electrodes and filled with a nematic treated as a resistor R. Typically $\tau_{c}$ is of the order of a few milliseconds. Therefore at the frequency used in experiments f=20mHz, which is much lower then $1/\tau_{c}$, the field \textbf{E} is produced in the so-called conductive regime. In the "one-gap" geometry the corresponding equipotentials are represented approximatively in Fig.\ref{fig:cond_diel}a. The electric field is uniform and its value is simply
\begin{equation}\label{eq:El_field}
E(t)=-U(t)/l_{g}
\end{equation}
At frequencies of the order of 10kHz, the field produced in the dielectric regime (see Fig.\ref{fig:cond_diel}b) would be much less uniform: its intensity would vary across the gap and would depend on the dielectric constants of the nematic and on dimensions of the cell.
\section{Measure of e$_{3}$-e$_{1}$ in MBBA and 5CB}
\label{sec:wound_up_texture}
\subsection{Geometry}
\label{sec:geometry}
In the plane/sphere geometry of our samples, the cuneitropism induces the radial configuration of the dowser field with \textbf{d}//\textbf{g}=\textbf{grad}h. A typical view of this texture with a $+2\pi$ defect (called \emph{residual monopole}) in its centre is shown in Figs.\ref{fig:MBBA_5CB}a and b. This radial dowser field \textbf{d} is submitted to the electric field E=-U/l$_{g}$ generated by a potential difference U between the ITO1 and ITO2 electrodes separated by the parallel gap of width l$_{g}$=1mm. As this relatively narrow gap is oriented radially, the electric field \textbf{E} is with a good approximation orthogonal to the initial dowser field \textbf{d}: $\psi_{e}\approx \pi/2$ (defined in Fig.\ref{fig:Tropisms}c).

This geometry has two advantages:
\begin{enumerate}
  \item the torque due to the induced polarisation given by Eq.\ref{eq:torque_pol_2} vanishes,
  \item the torque due to the flexo-electric polarisation is extremal: $\Gamma_{e}\approx -P_{fe}E$
\end{enumerate}
\subsection{Experimentum crucis, determination of the sign of e$_{3}$-e$_{1}$}
\label{sec:sign_e3_e1}
\begin{figure*}
\begin{center}
\includegraphics{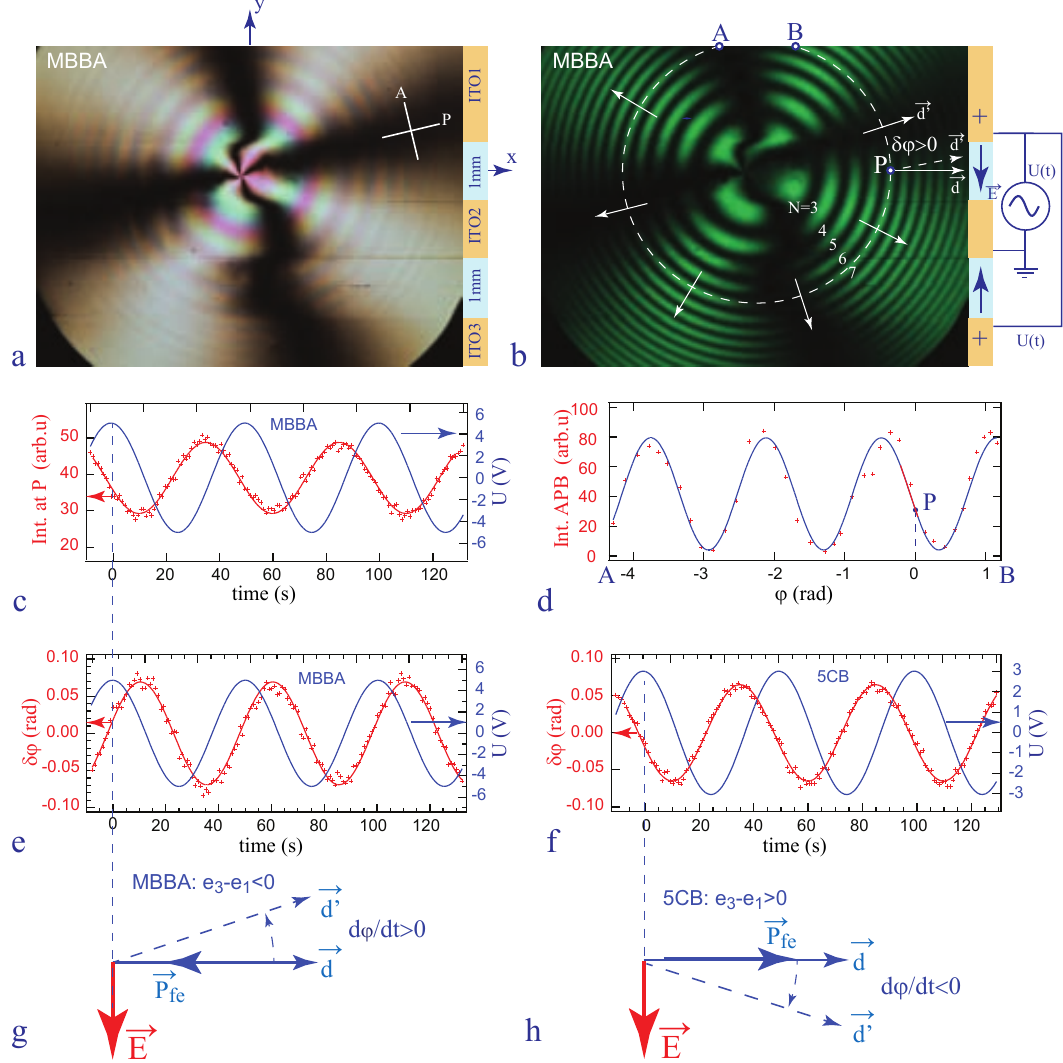}
\caption{Measure of the pertinent flexo-electric coefficient e$_{3}$-e$_{1}$ in MBBA and 5CB. a) View in white light of the equilibrium dowser texture of MBBA in its radial configuration induced by thickness gradients \textbf{grad}h. Crossed polarizers are slightly tilted with respect to the electric field \textbf{E} in the ITO1/ITO2 gap. b) The same view in monochromatic light. The system of concentric circular isochromes N=3, 4,... unveils the plane-sphere geometry of the sample. c) Plot of the transmitted light intensity I(t) measured in point P defined in b. d) Calibration of the I($\varphi$) relationship. The plot represents the light intensity measured along the circular dashed path APB defined in B. In the point P the slope is -141/rad. e) Using this value, the intensity plot I(t) shown in c has been transformed into the plot $\delta\varphi$(t) representing changes in the orientation of the dowser field in P induced by the alternating electric field E=U/l$_{g}$. Let's note that the plot $\delta\varphi$(t) is retarded by $\pi/2$ with respect to the applied field. f) Determination of the negative sign of e$_{3}$-e$_{1}$ in MBBA. g) Determination of the positive sign of e$_{3}$-e$_{1}$ in 5CB.}
\label{fig:MBBA_5CB}
\end{center}
\end{figure*}
Our first task was to determine the sign of the pertinent flexo-electric coefficient e$_{3}$-e$_{1}$ by means of a very simple \emph{experimentum crucis}. With the aim to detect directly rotation of the dowser field \textbf{d} induced by the electric field \textbf{E}, the crossed polarisers have been tilted by about $\pi/8$=22.5$^{\circ}$ with respect to the initial orientation \textbf{d}//\textbf{x} of the dowser field in the gap. In this geometry, variations $\delta\varphi$(t) of the orientation of the dowser field induced by the electric field E(t)=-U(t)/l$_{g}$  can be deduced from variations of the intensity of the transmitted light I(t).

Let us consider for example the point P located on the bright isochromatic interference fringe (marked in Fig.\ref{fig:MBBA_5CB}b with a dashed line). Upon application of the alternating voltage to the ITO1/ITO2 gap, the intensity of the transmitted light varies as shown in Fig.\ref{fig:MBBA_5CB}c. At the first maximum of the applied voltage, the intensity of the transmitted light I(t) is decreasing. This means that the dowser field is rotating in the anticlockwise direction out from the equilibrium orientation \textbf{d} toward \textbf{d'} (see Fig.\ref{fig:MBBA_5CB}b) for which the intensity of the transmitted light is lower.

For U=+5V, the electric field \textbf{E} has the -\textbf{y} direction as depicted in Fig.\ref{fig:MBBA_5CB}g. Thus the angle $\psi_{e}$ between \textbf{E} and \textbf{d} is increasing. This means that the direction of the torque exerted by the electric field on the dowser field is opposite to the one shown in Fig.\ref{fig:Tropisms}c. The flexo-electric polarisation \textbf{P}$_{fe}$ is thus opposite to \textbf{d} and P$_{fe}<$0.

Using the definition $\mathrm{P}_{fe}=(\pi/2)(e_{3}-e_{1})$ we can already conclude that in MBBA the flexo-electric coefficient e$_{3}$-e$_{1}$ is negative in agreement with the result of Dozov \emph{et al.} \cite{Dozov}.

With the aim to measure the value of e$_{3}$-e$_{1}$, one has to determine quantitatively variation $\delta\varphi$ of the dowser field angle. For this purpose we performed a calibration experiment which consists in measuring the intensity of the transmitted light I($\varphi$) along the circular dashed line path APB  defined in Fig.\ref{fig:MBBA_5CB}b. As expected, the plot of I($\varphi$) shown in Fig.\ref{fig:MBBA_5CB}d obeys to the expression:
\begin{equation}\label{eq:transmitted_light}
I(\varphi)=I_{o}\sin^2[2(\varphi-\alpha_{CP})]
\end{equation}
in which $\alpha_{CP}\approx\pi/8$ corresponds to the tilt angle of crossed polarizers with respect to \textbf{E}.

In the vicinity of the point P (where $\varphi$=0), one has a linear variation of I with $\varphi$
\begin{equation}\label{eq:transmitted_light_2}
I(\varphi)\approx 41.6-141\varphi
\end{equation}
with a negative slope dI/d$\varphi$=-141. Using this relationship between I and $\varphi$ we transformed the plot I(t) from Fig.\ref{fig:MBBA_5CB}c into the plot $\delta\varphi$(t) shown in Fig.\ref{fig:MBBA_5CB}e.

Before a detailed discussion of this plot obtained with MBBA, let us compare it with the analog plot shown in Fig.\ref{fig:MBBA_5CB}f obtained, by an identical experiment, with 5CB. Obviously, at the first maximum of the applied voltage (+3V instead of +5V in the case of MBBA), the angle $\psi_{e}$ between the dowser field and the electric decreases, contrary to the case of MBBA.

We can thus conclude that the sign of e$_{3}$-e$_{1}$ is positive in 5CB, contrary to the case of MBBA, and the flexo-electric polarisation \textbf{P}$_{fe}$ is parallel to \textbf{d} as shown in Fig.\ref{fig:Tropisms}c. This result is in disagreement with ref.\cite{Takezoe}.

\subsection{Determination of e$_{3}$-e$_{1}$}
\label{sec:measure_e3_e1}
From plots in Fig.\ref{fig:MBBA_5CB}e is obvious that the change $\delta\varphi$(t) in the orientation of the dowser field is retarded by $\pi/2$ with respect to the electric torque proportional to U(t)=U$_{o}$cos(2$\pi$ft). Such a dissipative behavior is expected when the driving electric torque is balanced mainly by the viscous torque:
\begin{equation}\label{eq:visc_motion}
-\gamma_{eff}h\frac{d\delta\varphi}{dt}-P_{fe}E_{o}\cos(2\pi ft)\approx 0
\end{equation}
Indeed, in such a case one has
\begin{equation}\label{eq:visc_motion_2}
\delta\varphi=\delta\varphi_{o}\sin(2\pi ft)
\end{equation}
with
\begin{equation}\label{eq:visc_motion_2}
\delta\varphi_{o}=-\frac{P_{fe}E_{o}}{\gamma_{eff}h\ 2\pi f}=\frac{P_{fe}U_{o}}{\gamma_{eff}h\ 2\pi f l_{g}}
\end{equation}
Using this expression we obtain an evaluation of the flexo-electric polarisation:
\begin{equation}\label{eq:pol_visc}
P_{fe}=\gamma_{eff}h\ 2\pi f l_{g}\frac{\delta\varphi_{o}}{U_{o}}
\end{equation}

From the plot in Fig.\ref{fig:MBBA_5CB}e we obtain $\delta\varphi_{o}$=0.07 for U$_{o}$=5V and f=20mHz. The thickness h in the point P can be estimated from the interference pattern inFig.\ref{fig:MBBA_5CB}b. As the point P is located on the bright fringe with the index N=6.5 we have
\begin{equation}\label{eq:h}
h=6.5\frac{\lambda}{\Delta n}
\end{equation}
with $\lambda\approx$ 0.53$\mu$m. The mean birefringence $\Delta n$ of the dowser texture is by definition:
\begin{equation}\label{eq:Delta_n_bis}
\Delta n=\frac{n_{o}}{h}\left[\int_{0}^{h}\frac{1}{\sqrt{1-C\sin^{2}\theta(z)}}-1\right]dz
\end{equation}
with
\begin{equation}\label{eq:R_bis}
C=\frac{n_{e}^{2}-n_{o}^{2}}{n_{e}^{2}}
\end{equation}
In MBBA  n$_{o}$=1.557 and n$_{e}$=1.792 \cite{Helfrich} so that C=0.245 and by numerical integration of the equation \ref{eq:Delta_n_bis} one obtains $\Delta n_{MBBA}$=0.111.  In the case of 5CB with n$_{o}$=1.51, n$_{e}$=1.75 and C=0.255 one obtains almost the same value of $\Delta n_{5CB}$=0.113. The local thickness in point P in Fig.\ref{fig:MBBA_5CB} is thus h=31$\mu$m in both cases of MBBA and 5CB.

The missing value of the effective rotation viscosity in equation \ref{eq:pol_visc} can be estimated as
\begin{equation}\label{eq:visc_gamma_eff}
\gamma_{eff}=\gamma_{1}\int_{0}^{\pi}\sin^2\theta d\theta=\gamma_{1}/2
\end{equation}
Using results of measurements of $\gamma_{1}$ made in MBBA and 5CB by Oswald et al. \cite{Oswald} one obtains $\gamma_{eff}$=50 mPa$^{.}$s.

Finally, using expression \ref{eq:pol_visc} we gets:
\begin{equation}\label{eq:pol_visc_MBBA}
P_{fe}\approx -2.7 \mathrm{pC/m} \ \mathrm{in}\ \mathrm{MBBA}
\end{equation}
and, using equation \ref{eq:mean_polarisation_2}
\begin{equation}\label{eq:e3_moins_e1_MBBA}
e_{3}-e_{1} \approx -1.7 \mathrm{pC/m} \ \mathrm{in}\ \mathrm{MBBA}
\end{equation}
In the case of 5CB (see Fig.\ref{fig:MBBA_5CB}f) we have measured $\delta\varphi_{o}$=-0.065 for U$_{o}$=3V and f=20mHz. Using h=31$\mu$m, $\gamma_{eff}$=50 mPa$^{.}$s we obtain
\begin{equation}\label{eq:pol_5CB}
P_{fe}\approx +4.2 \mathrm{pC/m} \ \mathrm{in}\ \mathrm{5CB}
\end{equation}
and
\begin{equation}\label{eq:e3_moins_e1_5CB}
e_{3}-e_{1} \approx +2.7 \mathrm{pC/m} \ \mathrm{in}\ \mathrm{5CB}
\end{equation}
\begin{figure*}
\begin{center}
\includegraphics{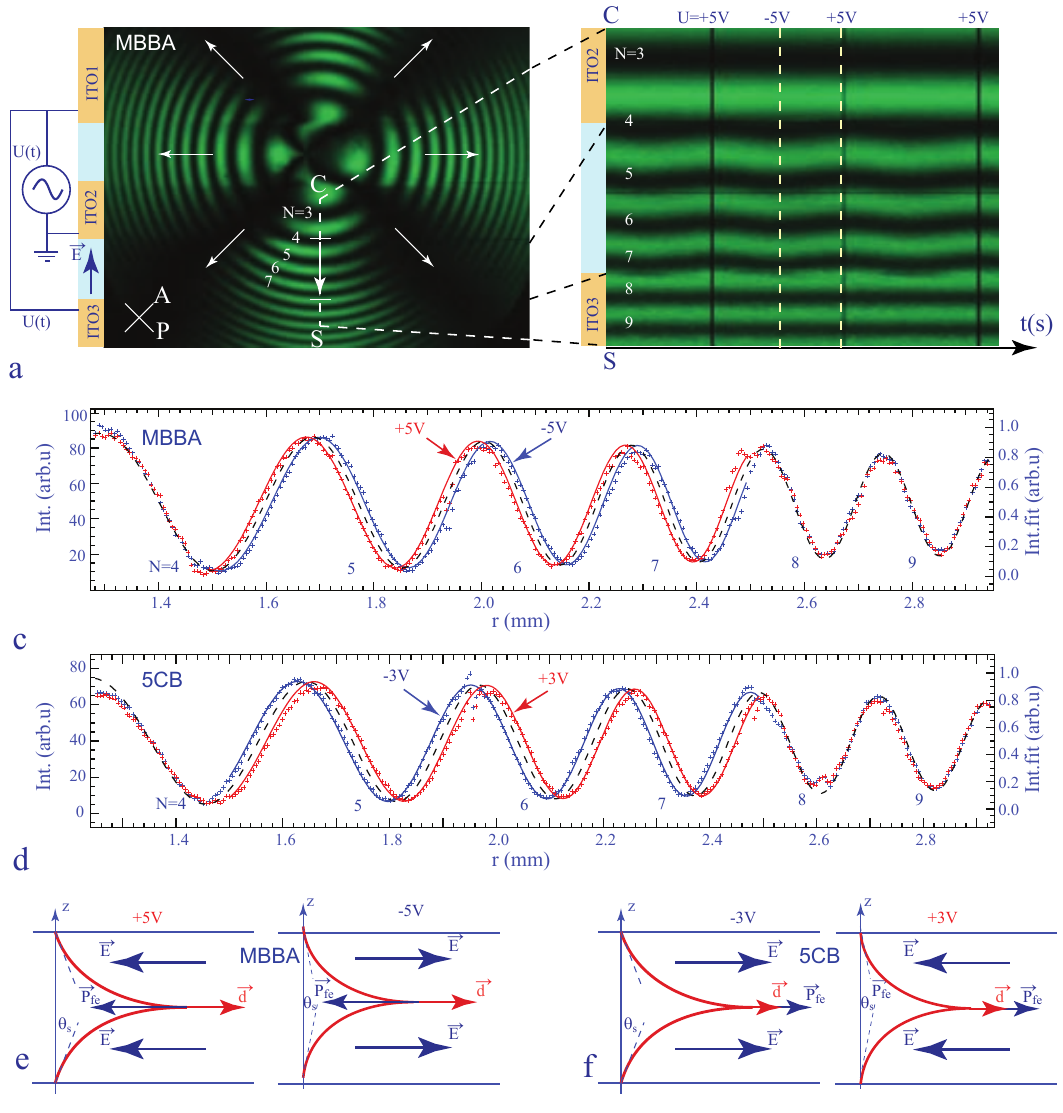}
\caption{Evidence for the flexo-electric modulation of the birefringence in the dowser texture. a) Interference pattern of the MBBA sample observed between crossed polarisers, tilted by $\pi$/4 with respect to the electric field in the ITO2/ITO3 gap. Along the dashed line CS, for U=+5V, the electric field \textbf{E} in the gap is opposite to the dowser field \textbf{d}. b) Spatio-temporal cross section extracted from a video along the line CS. Obviously, the N=5, 6 and 7 isochromes are set in motion by the electric field. c) Intensity profiles measured along the dashed lines labeled +5V and -5V in picture b. For U=+5V (-5V), the interference fringes are moving to the left (right) which means that the effective birefringence is increasing (decreasing). d) Result of the same experiment performed with 5CB. The motion of interference fringes is inversed: to the right for U=+5V and to the left for U=-5V. e-f) These modulation of the birefringence of the dowser texture are explained in terms of a weak hoemotropic anchoring. In MBBA, the polar angle $\theta_{s}$ increases for E$<$0 and decreases for E$>$0. In 5CB, the effect is inverse. }
\label{fig:Surf_effect}
\end{center}
\end{figure*}
\section{Flexo-electric modulation of birefringence}
\label{sec:modulation_Delta_n}
\subsection{Field-induced motion of isochromes in MBBA and 5CB}
\label{sec:modulation_Delta_n}
The interference figure in polarized light of the dowser texture in its radial configuration is made of four isogyres forming a maltese cross and of circular concentric isochromes unveiling thickness variation in the plane/sphere geometry. In experiments discussed above the flexo-electric polarisation was determined from the motion of the right isogyre making the angle of $\approx22.5^{\circ}$ with the ITO1/ITO2 gap.

During these experiments we observed unexpected motions of the N=5, 6 and 7 isochromes located in the second, ITO2/ITO3 gap. We have therefore performed another experiment with crossed polarizers tilted by $\pi/4$ with respect to the direction of the electric field as it is shown in Fig.\ref{fig:Surf_effect}a. In this configuration the contrast of isochromes on the dashed line CS is the best. As previously, an alternating voltage U(t)=U$_{o}$cos(2$\pi$ft) of frequency f=20mHz and amplitude U$_{o}$=5V was applied to the gap ITO2/ITO3 and a video with frame rate 1/sec was recorded.

Fig.\ref{fig:Surf_effect}b shows a spatio-temporal cross section of the video extracted along the dashed line CS defined in Fig.\ref{fig:Surf_effect}a. It is obvious that the N=5, 6 and 7 isochromes located in the ITO2/ITO3 gap are set in motion by the action of the electric field. To determine the phase relationship between this motion and the applied field we obturated briefly the camera at instants t$_{i}$ corresponding to the maximum of  the voltage U(t$_{i}$)=+5V. By this means we produced vertical black lines well visible in the spatio-temporal cross section. Clearly, there is no phase shift between the excitation U(t) and the displacement of isogyres: the response of the dowser texture to the applied field is purely elastic.

The amplitude of the displacement of isogyres has been determined from the intensity profiles I$_{-5V}$ and I$_{+5V}$ measured on the spatio-temporal cross section along dashed lines labeled -5V and +5V. Plotted in Fig.\ref{fig:Surf_effect}c, these profiles are shifted, with respect to their equilibrium positions, in sections corresponding to the N=5, 6 and 7 isochromes. The intensity plot labeled +5V is shifted to the left which corresponds to an increase of the mean birefringence $\Delta$n. The direction of the shift of the plot labeled -5V is opposite.

The mean birefringence $\Delta$n corresponding to these plots has been determined from fits to the expression:
\begin{equation}\label{eq:intensity_vs_r}
I(r)=I_{o}\sin^{2}(\delta/2)
\end{equation}
in which
\begin{equation}\label{eq:delta}
\delta=\frac{2\pi\Delta n}{\lambda}\left(h_{o}+R_{l}(1-\sqrt{1-(r/R_{l})^{2}}\right)
\end{equation}
The radius of curvature of the lens has been determined as R$_{l}$=107mm. The wavelength of the green monochromatic light is also known: $\lambda\approx$0.53$\mu$m. In the interval $1.5<r<2.5$ the best fits of experimental points labeled +5V and -5V were obtained with h$_{o}$=22.4$\mu$m and $\Delta$n$_{+5V}$=0.097 and $\Delta$n$_{-5V}$=0.096. Outside of this interval, for the equilibrium dowser texture, the best fit gave $\Delta$n$_{o}$=0.0965.

We repeated the same experiment with 5CB. The corresponding intensity profiles are plotted in Fig.\ref{fig:Surf_effect}d. It is obvious that the field-induced shifts are opposite here: the profiles labeled +3V and -3V are shifted respectively to the right and to left. Using R$_{l}$=107mm and h$_{o}$=22.9$\mu$m, we obtained : $\Delta$n$_{+3V}$=0.0964, $\Delta$n$_{-3V}$=0.0977 and $\Delta$n$_{o}$=0.097.

We can conclude that upon the application of an alternating potential difference U(t)=U$_{o}$cos(2$\pi$ft) the mean birefringence of the dower texture varies as:
\begin{equation}\label{eq:Delta_n_o}
\Delta n(t)\approx\Delta n_{o}(1+\epsilon\cos(2\pi ft))
\end{equation}
with $\epsilon$=+0.005 for U$_{o}$=5V in MBBA and $\epsilon$=-0.006 for U$_{o}$=3V in 5CB.
\subsection{Field-induced surface effect}
\label{sec:Surface_effect}
Experimental results described above are similar to those obtained by Madhusudana and Durand \cite{Madhusudana} in cells with hybrid anchorings when the electric field E is parallel to the z axis instead the y axis as in experiments of Dozov \emph{at al.} (see Fig.\ref{fig:Flexo_pol}b). They have interpreted them in terms of a model taking into account a finite strength of the homeotropic anchoring on one of the limit glass slides. The planar anchoring on the other slide was supposed to be infinitely strong.

In the present case of the dowser texture this last assumption is not necessary for the anchorings on both limit surfaces are homeotropic and of the same strength.

Below, we will adapt the Madhusudana and Durand's model to the case of the dowser texture. To start with, we will use one elastic constant approximation for which the energy per unit area of the dowser texture can be expressed as
\begin{equation}\label{eq:surf_energy}
F_{s}=\int_{0}^{h}\left[\frac{K}{2}\left(\frac{d\theta}{dz}\right)^{2}-P_{fex}E\right]dz+2\frac{W}{2}\sin^{2}\theta_{s}
\end{equation}
with
\begin{equation}\label{eq:surf_Pefx}
P_{fex}=(e_{3}\cos^{2}\theta-e_{1}\sin^{2}\theta)\frac{d\theta}{dz}
\end{equation}
The last term in equation \ref{eq:surf_energy} corresponds to the sum of anchoring energies on the glass surfaces with homeotropic anchoring.

Minimisation of F$_{s}$ leads to the Euler-Lagrange equation
\begin{equation}\label{eq:surf_E_L}
\frac{d^{2}\theta}{dz^{2}}=0
\end{equation}
with the solution:
\begin{equation}\label{eq:theta_of_z}
\theta(z)=\theta_{s}+\frac{(\pi-2\theta_{s})z}{h}
\end{equation}
satisfying boundary conditions:
\begin{equation}\label{eq:boundary_cond}
\theta(0)=\theta_{s}\ \ \mathrm{and} \ \ \theta(h)=\pi-\theta_{s}
\end{equation}
Inserting this solution to the expression of energy \ref{eq:surf_energy}, one obtains:
\begin{equation}\label{eq_surf_en_2}
F_{s}=\frac{K}{2}\frac{(\pi-2\theta_{s})^{2}}{h}-E\left[\frac{(e_{3}-e_{1})(\pi-2\theta_{s})}{2}-\frac{(e_{1}+e_{3})\sin2\theta_{s}}{2}\right]+\\ W\sin^{2}\theta_{s}
\end{equation}
Minimisation of F$_{s}$ with respect to $\theta_{s}$ leads to
\begin{equation}\label{surf_torques}
-K\frac{\pi-2\theta_{s}}{h}+E\left[(e_{3}-e_{1})+(e_{1}+e_{3})\cos2\theta_{s}\right]+2W\sin\theta_{s}\cos\theta_{s}=0
\end{equation}
or, in the limit of small $\theta_{s}$, to
\begin{equation}\label{theta_s}
\left(1+\frac{K/W}{h}\right)\theta_{s}\approx \frac{\pi}{2}\frac{K/W}{h}-\frac{Ee_{3}}{W}
\end{equation}
When the ratio K/W=L known as extrapolation length is much smaller than h, one obtains a very simple expression:
\begin{equation}\label{theta_s_2}
\theta_{s}\approx \frac{\pi}{2}\frac{K/W}{h}-\frac{Ee_{3}}{W}
\end{equation}
In the absence of the electric field, for E=0, the angle $\theta_{s}$ of the tilt of the homeotropic anchoring is proportional to the elastic torque K(($\pi$/2)/h) and inversely proportional to the anchoring strength W. The application of the electric field changes is it by $\delta\theta_{s}$=-Ee$_{3}$/W.

\subsection{Determination of e$_{3}$}
\label{sec:e_3}

We have now to calculate changes in the mean birefringence due the field-driven modulation of the anchoring angle $\theta_{s}$.
Inserting eq.\ref{eq:theta_of_z} in \ref{eq:Delta_n_bis} one obtains for small C ($\approx$0.24)
\begin{equation}\label{eq:Delta_n_3}
\Delta n\approx\frac{n_{o}C}{4}\left[1+\frac{\theta_{s}}{\pi/2}\right]
\end{equation}
or, using expressions \ref{theta_s_2} and \ref{eq:El_field}
\begin{equation}\label{eq:Delta_n_4}
\Delta n\approx\Delta n_{o}\left[1+\frac{L}{h}+\frac{(e_{3}U/Wl_{g})}{\pi/2}\right]
\end{equation}
with
\begin{equation}\label{eq:Delta_n_5}
\Delta n_{o}=\frac{n_{o}C}{4}
\end{equation}
As the equation \ref{eq:Delta_n_4} corresponds to the equation \ref{eq:Delta_n_o}, we obtain
\begin{equation}\label{eq:epsilon_e3}
\epsilon=\frac{(e_{3}U/Wl_{g})}{\pi/2}
\end{equation}
\begin{figure}
\begin{center}
\includegraphics{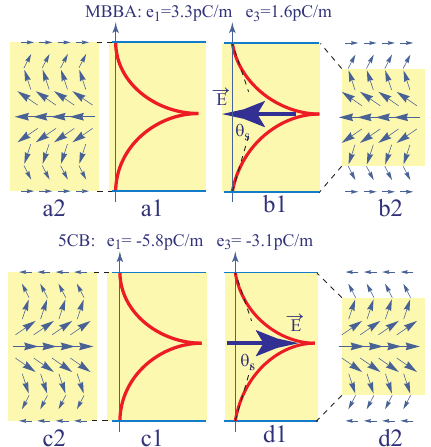}\\
\caption{Interpretation of the observed field-induced modulation of the birefringence of the dowser texture. a-b) MBBA, c-d) 5CB. a) The dowser texture (a1) and its flexo-electric polarisation density (a2) for \textbf{E}=0. b) Dowser texture (b1) deformed by application of the field \textbf{E}. The corresponding polarisation density (b2) is obtained from a2 by truncation of bands with polarisation density opposite to \textbf{E}. c) The dowser texture (c1) and its flexo-electric polarisation density (c2) for \textbf{E}=0. b) Dowser texture (d1) deformed by application of the field \textbf{E}. The corresponding polarisation density (d2) is obtained from c2 by truncation of bands with polarisation density opposite to \textbf{E}. }
\label{fig:explication}
\end{center}
\end{figure}
\begin{figure*}
\begin{center}
\includegraphics{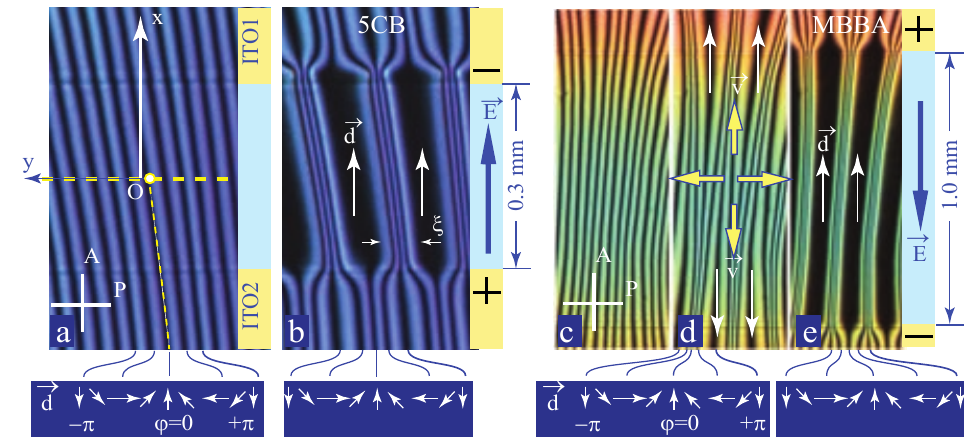}
\caption{Second evidence of the flexo-electric polarisation of the dowser texture in 5CB (a-b) and MBBA (c-e). a) System of quasi-equidistant isogyres in a wound up dowser texture of 5CB, E=0. b) Formation of 2$\pi$ walls in electric field. For 5CB, the dowser field \textbf{d} in the enlarged isogyres has the direction of the electric field \textbf{E}. c) System of quasi-equidistant isogyres in a wound up dowser texture of MBBA, E=0. d) Determination of the dowser field from the action of a divergent radial flow induced by a decrease in the lens/slide distance h$_{o}$. The dowser field \textbf{d} in enlarged isogyres is aligned in the direction of the flow \textbf{v}. e) Formation of 2$\pi$ walls in electric field. For MBBA, the dowser field \textbf{d} in the enlarged isogyres has the direction opposite to the electric field \textbf{E}.}
\label{fig:Walls_defects}
\end{center}
\end{figure*}
To determine the value of e$_{3}$ from this expression we will use results of the experiment discussed in section \ref{sec:modulation_Delta_n}: $\epsilon$=0.005 for U$_{o}$=5V for MBBA and $\epsilon$=-0.006 for U$_{o}$=3V in 5CB.
The homeotropic anchoring obtained with the egg yolk being weak, we will estimate it strength as W$\approx$10$^{-6}$J/m$^{2}$ (see ref.\cite{Jerome}). Finally, with l$_{g}$=1mm, we obtain
\begin{equation}\label{eq:epsilon_e3_MBBA}
e_{3}=+1.6pC/m\  \  \mathrm{for \  \  MBBA}
\end{equation}
and
\begin{equation}\label{eq:epsilon_e3_5CB}
e_{3}=-3.1pC/m\  \  \mathrm{for \  \  5CB}
\end{equation}

\subsection{Determination of e$_{1}$}
\label{sec:e_1}
Using values of e$_{3}$-e$_{1}$ obtained in section \ref{sec:measure_e3_e1} we can finally calculate values of e$_{1}$:
\begin{equation}\label{eq:epsilon_e1_MBBA}
e_{1}=+3.3pC/m\  \  \mathrm{for \  \  MBBA}
\end{equation}
and
\begin{equation}\label{eq:epsilon_e1_5CB}
e_{1}=-5.8pC/m\  \  \mathrm{for \  \  5CB}
\end{equation}
\section{Discussion}
\subsection{Qualitative explanation of experimental results}
\label{sec:qual_expl}
Using these values of the flexo-electric coefficients determined in our experiments we can select two polarisation patterns from the set of 16 typical patterns shown in Fig.\ref{fig:polar_e1_e3}. They are surrounded by dashed-line frames in Fig.\ref{fig:polar_e1_e3} and labeled MBBA and 5CB. For convenience sake they are reproduced in Fig.\ref{fig:explication}.

Let us consider first the case of MBBA. Its polarisation pattern is shown in Fig.\ref{fig:explication}a2 beside the unperturbed dowser texture (a1). Upon application of the electric field in the direction opposite to the dowser field, the polar angle $\theta_{s}$ grows and the dowser texture is deformed as shown in b1. One can see it as a truncated version of the unperturbed one. For this reason, the corresponding polarisation pattern shown in b2 is a truncated version of the equilibrium one shown in a2. Obviously, by removing bands with polarisation \textbf{P}$_{fe}$ opposite to \textbf{E}, the energy -\textbf{P}$\cdot$\textbf{E} is lowered.

In the case of 5CB (Figs.\ref{fig:explication}c and d), the same arguments held when the field \textbf{E} is applied in the direction of the dowser field.
\subsection{Perspectives}
Once the existence of the flexo-electric polarisation in the radial dowser texture is firmly established and approximate values of the flexo-electric coefficients in MBBA and 5CB are known, we could discuss results of experiments with more complex dowser fields, possibly containing +2$\pi$ and -2$\pi$ defects, submitted to electric fields with more complex geometries. As these results are too abundant for their presentation here we postpone their discussion to other forthcoming papers. Nevertheless, it seems useful to mention here briefly two of them.
\subsubsection{Wound up dowser texture}
\label{sec:walls_wound_up}
The first one is related to the work of Link \emph{et al.} \cite{Takezoe} on the width $\xi$ of 2$\pi$ walls created be an electric field in a wound up \textbf{c} field of a 5CB layer floating on glycerol. In the wound up \textbf{c} field the azimuthal angle $\varphi$ varies, let us say, as $\varphi(y)$=2$\pi$y/$\Lambda$. When observed  between crossed polarizers (parallel to x and y axes), such a wound up \textbf{c} texture appears as a system of equidistant black isogyres parallel to the x axis and located at y$_{i}$=i$\Lambda$/4, with integer i. Upon application of the electric field in \textbf{x} direction, Link et al. observed that isogyres corresponding to \textbf{c}//\textbf{E}, are enlarged while the other isogyres are assembled into 2$\pi$ walls. Using the definition of the \textbf{c} vector given in ref.\cite{Takezoe} and reproduced here in Fig.\ref{fig:Flexo_pol}c, the authors concluded that the flexo-electric polarisation \textbf{P} has the direction of the \textbf{c} field so that the pertinent coefficient e$_{3}$-e$_{1}$ is negative. From the width $\xi$ of 2$\pi$ walls Link et al. determined e$_{3}$-e$_{1}$=-11pC/m.

We have performed similar experiments with a dowser texture \textbf{d} wound up in y direction by the method described in ref.\cite{PP_Hulin_MHG_rheo}. In the case of 5CB, upon application of the electric field in \textbf{x} direction, the system of equidistant black isogyres (see Fig.\ref{fig:Walls_defects}a) was split into 2$\pi$ walls separated by bands in which the dowser field \textbf{d} is parallel to the applied field \textbf{E} (see Fig.\ref{fig:Walls_defects}b). This means that in 5CB the flexo-electric polarisation \textbf{P} has the direction of the \textbf{d} field.

In the case of MBBA (see Figs.\ref{fig:Walls_defects}c, d and e) when the electric field was applied in -\textbf{x} direction the system of equidistant isogyres shown in Fig.\ref{fig:Walls_defects}c was split into 2$\pi$ separated by bands in which the dowser field \textbf{d} is antiparallel to the applied field \textbf{E} (see Fig.\ref{fig:Walls_defects}e). This means that in MBBA the flexo-electric polarisation \textbf{P} has the direction oppposite to the \textbf{d} field.

These two experiments with a wound up dowser texture lead to the same conclusions as those of experiments with the radial dowser texture discussed in section \ref{sec:wound_up_texture}: in 5CB the pertinent flexo-electric coefficient e$_{3}$-e$_{1}$ is positive while in MBBA
it is negative.

Let us emphasize that conclusions of experiments with the wound up dowser texture depend crucially on the detection of the dowser field direction in isogyres enlarged by application of the electric field. To determine this direction we submitted the wound up dowser texture to the action of a divergent radial flow driven by decreasing the distance h$_{o}$ between the lens and the glass slide. The pattern of initially equidistant isogyres was perturbed as shown in Fig.\ref{fig:Walls_defects}d: isogyres in which the dowser field \textbf{d} has the same direction as the flow velocity \textbf{v} are enlarged by rheotropism discussed recently in ref.\cite{PP_Hulin_MHG_rheo}.
\begin{figure}
\begin{center}
\includegraphics{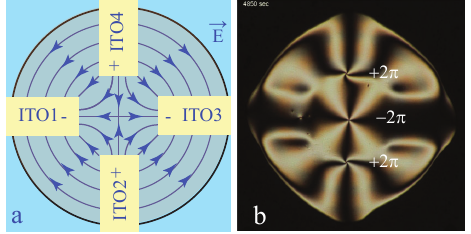}
\caption{Manipulation of defects in the dowser texture by an electric field. a) Geometry of the field generated by a system of four ITO electrodes. b) Stable configuration of three defects.  }
\label{fig:Manipulation_defects}
\end{center}
\end{figure}
\subsubsection{Manipulation of defects}
The second experiment depicted in Fig.\ref{fig:Manipulation_defects} proves that as expected the defects of the dowser texture can be manipulated by electric fields of appropriate geometry. In Fig.\ref{fig:Manipulation_defects}b, as expected by analogy with the work of Giomi \emph{et al.} \cite{Giomi_Sengupta}, a -2$\pi$ defect is trapped in the stagnation point of a quadrupolar electric field generated by a system of four ITO electrodes. Let us emphasize that beside this -2$\pi$ defect the dowser field inside the nematic droplet contains, for topological reason, also two +2$\pi$ defects. Their positions are also stable.

\section*{Conflicts of interest}
There are no conflicts to declare.

\section*{Acknowledgements}
P.P. is grateful to L. Giomi, T. Lubensky and V. Vitelli for the invitation to the Lorentz Workshop on Topology in Complex Fluids and for fruitful discussions about generation and handling of topological defects. We thank P. Oswald for discussions and for the information about the reference \cite{Oswald} containing careful measurements of the viscosity coefficient $\gamma_{1}$. This work benefited from the technical assistance of V. Klein, S. Saranga, J. Sanchez, M. Bottineau, J. Vieira, I. Nimaga and C. Goldmann. P.P. thanks also S. Ravy for a financial support. MHG thanks FEDER funds through the COMPETE 2020 Program and National Funds through FCT-Portuguese Foundation for Science and Technology under projects numbers POCI-01-0145- FEDER-007688 (Reference UID/CTM/50025).

\end{document}